\documentclass[10]{article}
\usepackage{arxiv}
\usepackage[utf8]{inputenc}
\usepackage[T1]{fontenc}
\usepackage{xcolor}
\usepackage[round]{natbib}
\definecolor{blendedblue}{rgb}{0.2, 0.2, 0.6}
\usepackage{graphicx}
\usepackage[bookmarks   = true,
            citecolor   = blendedblue,
            colorlinks  = true,
            linkcolor   = blendedblue,
            urlcolor    = blendedblue,
            citecolor   = blendedblue,
            linktocpage = false,
            hyperindex  = true]{hyperref}
\usepackage{booktabs}
\usepackage{amsfonts}
\usepackage{nicefrac}
\usepackage{amsmath, amssymb, amsthm}
\usepackage{url}
\usepackage{doi}
\usepackage{bm}
\usepackage{fontawesome}
\usepackage{enumerate}
\usepackage{mathtools}
\usepackage{lmodern}
\usepackage{siunitx}
\usepackage{etoolbox}
\usepackage{calc}
\usepackage{dsfont}
\usepackage{multirow}
\usepackage{array}
\graphicspath{{figures/}}
\usepackage{todonotes}
\usepackage{changes}
\usepackage{enumitem} 
\usepackage{orcidlink}

\usepackage[normalem]{ulem}
\usepackage[labelsep=period]{caption}

\renewcommand{\headeright}{}

\usepackage{parskip}
\makeatletter
\renewcommand\@biblabel[1]{#1.}
\makeatother

\title{Covariate-Dependent Functional Principal Component Analysis for SHM}
\author{
    Philipp Wittenberg \orcidlink{0000-0001-7151-8243} \\
        Chair of Statistics and Data Science \\
 	Dept.~of Mathematics and Statistics\\
	School of Business, Economics and Social Sciences\\
        Helmut Schmidt University\\
	Hamburg, Germany\\
	\texttt{pwitten@hsu-hh.de} \\
    \And
    Lizzie Neumann \orcidlink{0000-0003-2256-1127} \\
        Chair of Statistics and Data Science \\
 	Dept.~of Mathematics and Statistics\\
	School of Business, Economics and Social Sciences\\
        Helmut Schmidt University\\
	Hamburg, Germany\\
	\texttt{neumannl@hsu-hh.de} \\
 	\And
    Kristof Maes \orcidlink{0000-0003-4188-3180} \\
        Structural Mechanics Section\\
 	Dept.~of Civil Engineering\\
	Faculty of Engineering Science\\
        KU Leuven\\
	Leuven, Belgium\\
	\texttt{kristof.maes@kuleuven.be} \\
 	\And
    Jan Gertheiss \orcidlink{0000-0001-6777-4746} \\
        Chair of Statistics and Data Science \\
 	Dept.~of Mathematics and Statistics\\
 	School of Business, Economics and Social Sciences\\
        Helmut Schmidt University\\
	Hamburg, Germany\\
	\texttt{gertheij@hsu-hh.de} \\
}
\begin{document}

\maketitle

\begin{abstract}
In Structural Health Monitoring (SHM), sensor measurements and derived features such as eigenfrequencies often exhibit systematic daily patterns and can therefore be naturally represented as functional data. Furthermore, these patterns are typically influenced by environmental factors, particularly temperature, which can substantially affect the observed system response. While most existing methods for removing environmental effects assume that confounding influences affect only the mean response, it has been shown that environmental and operational factors may also alter the covariance structure of the residual process. To address this limitation in a functional data monitoring framework, we incorporate so-called covariate-dependent functional principal component analysis (CD-FPCA), which allows eigenfunctions and eigenvalues of the residual process to vary smoothly with covariates such as temperature. The proposed methodology is illustrated using an extended version of the KW51 railway bridge eigenfrequency dataset. This case study suggests that accounting for covariate effects beyond the functional mean can improve the robustness of the monitoring procedure, in particular by reducing environmentally induced (false) alarms under challenging low-temperature conditions.
\end{abstract}
\bigskip
\noindent%

\section{Introduction and Data}\label{sec:intro}
In structural health monitoring (SHM), it is well known that the system outputs, i.e., sensor measurements and derived damage-sensitive features (such as natural frequencies), change not only due to structural damage but also in response to environmental and operational variations (EOV), particularly due to temperature \citep{Han.etal_2021, Wang.etal_2022}. Additionally, system outputs often exhibit systematic, recurring daily patterns, making them well-suited to a functional data representation \citep{Wittenberg.etal_2026}. To account for environmental variability from a functional perspective, \cite{Wittenberg.etal_2025a} recently proposed the covariate-adjusted functional data analysis framework for SHM (CAFDA-SHM), which combines nonlinear function-on-function regression with functional principal component analysis (FPCA) applied to the residual process.
However, as with most existing methods for removing environmental effects, the CAFDA framework assumes that EOVs affect only the mean value of the system output. Recent works by \cite{Neumann.etal_2025a, Neumann.etal_2025b, Neumann.etal_2025c}, however, demonstrated that EOVs may also affect higher-order moments, particularly the output's (co-)variance structure. That is why incorporating conditional (co-)variance information into the monitoring framework could reduce the risk of false alarms and increase the probability of detecting actual damage.

\begin{figure}[h]
    \centering
    \includegraphics[width=.9\linewidth]{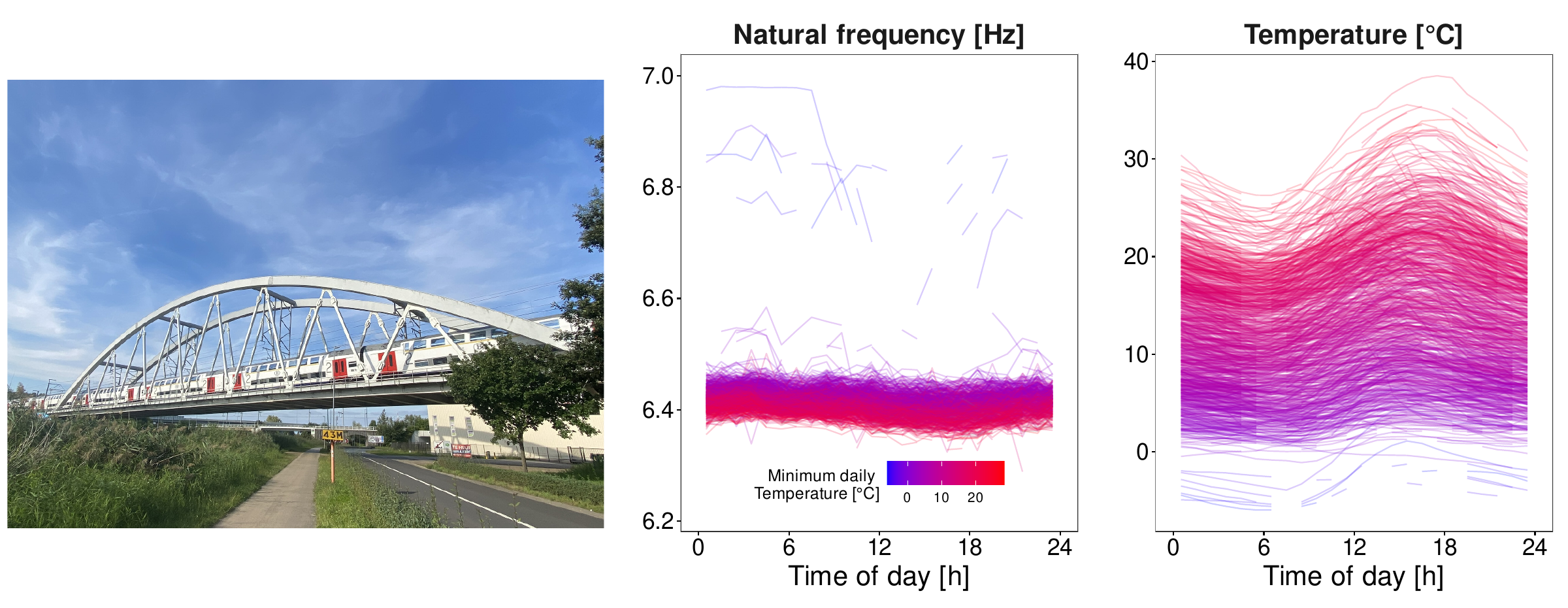}
    \caption{South view on bridge KW51 (left), considered profiles of natural frequencies (Mode 13, middle) and bridge surface temperature (right).}
    \label{fig:data}
\end{figure}

To illustrate both the functional perspective and the complex nature of environmental effects, we will consider the natural frequency dataset obtained from the long-term monitoring of railway bridge KW51 in Leuven, Belgium \citep{Maes.Lombaert_2020, Maes.Lombaert_2021}, see Figure \ref{fig:data}~(left). 
The evolution of natural frequencies from 1st October 2018 to 30th September 2022 shows a significant temperature influence across most modes, particularly during frost periods. Apart from those, there was also a large influence of the retrofitting works~\citep{Maes.etal_2022}. In this paper, we focus on the structure's response during the period following the retrofit, i.e., from 27th May 2019 to 30th September 2022. Note that this is a substantial extension of the original KW51 dataset presented in~\cite{Maes.Lombaert_2020, Maes.Lombaert_2021}. 

Figure \ref{fig:data} (middle) displays the data in functional data format, i.e., the natural frequency (Mode 13) is considered as a function of the time of day (0h to 24h). The tracked natural frequencies, however, contain substantial missing data due to the fact that not all modes were sufficiently excited to enable a 100\% successful identification~\citep{Maes.Lombaert_2021}. We selected Mode 13 (the fifth-order vertical bending mode) because it shows a clear increase in natural frequency at cold temperatures, which is attributed to frost in the ballast layer. The fact that the data stream is characterized by gaps, together with highly nonlinear behavior in response to changing temperatures, poses another challenge for the modeling procedure. To address all these issues, this article extends the CAFDA framework by integrating a covariate-dependent version of functional principal component analysis (CD-FPCA) developed by \cite{Jiang.Wang_2010}. CD-FPCA allows both the mean and covariance structure to depend on environmental covariates, thereby potentially improving robustness to environmental effects in SHM applications. The remainder of the article is structured as follows. Section~\ref{sec:methods} presents CD-FPCA and its integration with CAFDA-SHM. Section~\ref{sec:results} reports the results on the KW51 data. Section~\ref{sec:discussion_conclusion} concludes the study and outlines potential extensions.
\section{Methodology}\label{sec:methods}
\subsection{Covariate-dependent FPCA}\label{subsec:CD-FPCA}
To address the challenges outlined in Section \ref{sec:intro}, we adapt the CD-FPCA approach developed by \cite{Jiang.Wang_2010}. This approach consists of two components: a systematic part considering the mean function and a stochastic part comprising random components that reflect the covariance structure. The former, as discussed in \cite{Wittenberg.etal_2025a, Wittenberg.etal_2025b}, can be modeled using functional regression that incorporates the effects of covariates on the mean. The latter, unlike classical FPCA, explicitly includes a covariance structure that depends on covariates. The original CD-FPCA works as follows. The covariate $z$ may be a one-dimensional, scalar quantity like the minimum daily temperature as used for coloring in Figure~\ref{fig:data}~(middle/right), or multivariate, vector-valued; however, due to the so-called ``curse of dimensionality'', only a low-dimensional covariate is practically feasible. To keep things simple, we will focus on a one-dimensional $z$. Then, it is assumed that the quantity of interest $y(t,z)$, here, the natural frequency $y$ at time $t \in \mathcal{T}$ for a given $z$, has the form 
\begin{equation}\label{eq:initial_model}
    y(t,z)=\mu(t,z)+\sum_{r=1}^\infty\xi_r\phi_r(t,z),
\end{equation}
where $\mu(t,z)$ is the (conditional) mean of $y$ at time $t$ for a given $z$; $\phi_r(t,z)$, $r=1,2,\ldots$, are the eigenfunctions of the (conditional) covariance operator $\Gamma(s,t,z) = \text{Cov}(y(s,z),y(t,z))$, and $\xi_r$ are uncorrelated random scores with mean 0 and variance $\nu_r(z)$, with $\nu_1(z)\geq\nu_2(z)\geq\dots\geq 0$ being the eigenvalues of $\Gamma(s,t,z)$. The (observable) system output $u_{ij}=y(t_{ij}, z_i) + \epsilon_{ij}$
represents the $j$th observation on day $i$ of the (underlying) $y$, recorded at a time point $t_{ij}\in \mathcal{T}$, $\mathcal{T}=(0h,24h]$, $i=1,\dots,n$, and $j=1,\dots,n_i$. Each observation has an associated covariate $z_i\in\mathbb{R}$ (here, minimum daily temperature) and additionally, an independent and identically distributed measurement error/white noise $\epsilon_{ij}$ with mean zero and constant variance $\sigma^2$. Typically, the sum in~\eqref{eq:initial_model} is truncated at some value $m$ resulting in the model
\begin{equation}\label{eq:observed_data}
    u_{ij}=\mu(t_{ij}, z_i) + \sum_{r=1}^m\xi_{ir}\phi_r(t_{ij},z_i) + \epsilon_{ij},
\end{equation}
with scores $\xi_{ir}$ and measurement errors $\epsilon_{ij}$ being mutually independent.

In a real-world SHM scenario, however, the temperature is not only observed in terms of its daily minimum (or other aggregated values), but as a profile over the entire day (compare Figure~\ref{fig:data}, right), and the (expected) natural frequency at time~$t$ should rather depend on the material temperature $x(t)$ at the very time $t$ than the daily minimum temperature $z$. Following \cite{Wittenberg.etal_2025a}, we hence replace $\mu(t, z)$ in~\eqref{eq:initial_model} by $\mu(t,x(t)) = \alpha_0 + \tilde{\alpha}(t) + f(x(t))$ and $\mu(t_{ij}, z_i)$ in~\eqref{eq:observed_data} by $\mu(t_{ij},x_i(t_{ij})) = \alpha_0 + \tilde{\alpha}(t_{ij}) + f(x_i(t_{ij}))$, respectively, where $f$ describes the, concurrent and potentially nonlinear, effect of the (material) temperature. The overall constant $\alpha_0$ together with the so-called (centered) functional intercept $\tilde{\alpha}$ model (potential) recurring daily patterns that are, e.g., induced by other (unobserved) covariates and cannot be explained through the temperature alone. Concerning the covariance operator $\Gamma(s,t,z)$, it is still assumed that the aggregated $z$ is sufficient to account for the higher-order effects beyond the mean $\mu$. A key step in the corresponding CD-FPCA procedure is then the estimation of the covariate-dependent covariance function $\Gamma(s,t,z)$, from which covariate-specific eigenvalues and eigenfunctions follow via the associated eigendecomposition. The covariance estimator is obtained by smoothing the raw covariances
\begin{equation}\label{eq:residuals}
    c_{ijk}=\big(u_{ij}-\hat{\mu}(t_{ij}, x_i(t_{ij}))\big)\big(u_{ik}-\hat{\mu}(t_{ik}, x_i(t_{ik}))\big),
\end{equation}
where the mean function $\mu$ is estimated in the framework of functional additive mixed models~\citep{Ivanescu.etal_2015, Scheipl.etal_2016, Greven.Scheipl_2017} as described in \cite{Wittenberg.etal_2025a}.
For a one-dimensional $z$, estimating $\Gamma(t,s,z)$ requires smoothing a three-dimensional surface of $c_{ijk}$ over $(t_{ij}$,$t_{ik}, z_i)$ for all $j,k=1,\dots,n_i,$ $i=1,\dots,n$. As in \cite{Jiang.Wang_2010}, we employ a local-linear smoother in terms of
\begin{equation}\label{eq:cov_func_loc_lin}
\begin{split}
\hat{\Gamma}(t,s,z)=\hat{\beta}_0,\quad\text{where} \\
    \boldsymbol{\hat{\beta}}  = \underset{\boldsymbol{\beta} = (\beta_0,\beta_1,\beta_2,\beta_3)^\top}{\arg\min}\Biggl\{
    \sum_{i=1}^n\sum_{1\leq j\neq k\leq n_i} & K_3\left(\frac{t-t_{ij}}{h_t}, \frac{s-t_{ik}}{h_t}, \frac{z-z_{i}}{h_z}\right) \\
    & \times \big[c_{ijk}-\big(\beta_0+\beta_1(t_{ij}-t) + \beta_2(t_{ik}-s)+\beta_3(z_i-z)\big)\big]^2\Biggl\},
\end{split}
\end{equation}
with $K_3$ being a three-dimensional kernel function. The bandwidths $h_t, h_z$ can be selected via $k$-fold cross-validation; see, e.g., \cite{Petersen.etal_2019}.

\subsection{Estimation and monitoring of the scores}\label{subsec:estimation_scores}
Assuming normally distributed principal component scores and measurement errors, the scores can be estimated following the approach of \cite{Yao.etal_2005}
\begin{equation}\label{eq:generate_scores}
    \hat{\xi}_{ir} = \hat{\nu}_r(z_i)\hat{\boldsymbol{\phi}}^\top_{ir}\hat{\boldsymbol{\Sigma}}_i^{-1}(\boldsymbol{u}_i-\hat{\boldsymbol{\mu}}_i) \quad r=1,\dots,m,
\end{equation}
 where $\boldsymbol{u}_i = (u_{i1},\ldots,u_{i,n_i})^\top$, $\hat{\boldsymbol{\mu}}_i=(\hat{\mu}(t_{i1}, x_i(t_{i1})), \dots, \hat{\mu}(t_{i,n_i}, x_i(t_{i,n_i})))^\top$, $\hat{\boldsymbol{\phi}}_{ir}=(\hat{\phi}_{r}(t_{i1}, z_i), \dots$ $\dots, \hat{\phi}_{r}(t_{i,n_i}, z_i))^\top$,  $(\hat{\boldsymbol{\Sigma}}_i)_{j,k}=\hat{\Gamma}(t_{ij},t_{ik}, z_i) + \hat{\sigma}^2\delta_{jk}$, and $\delta_{jk} = 1$ if $j=k$ and 0 otherwise. The covariate-dependent eigensystem $\{\hat{\phi}_r(\cdot,z), \hat{\nu}_r(z)\}$, mean function $\mu$, and error variance $\sigma^2$ are estimated from Phase~I/training data and, in theory, both $\hat{\phi}_r(\cdot,z)$ and $\hat{\nu}_r(z)$ are smooth functions of $z$. In practice, to save computational time during monitoring, they are already evaluated over the relevant range of the covariate on a reasonably fine grid $z^{(1)},\ldots,z^{(G)}$ during training. In the actual monitoring phase, so-called Phase~II, for a new day $l=1,2,\ldots$ with covariate $z_l$, we select the nearest grid point $z^{(g(l))}$ and use the corresponding eigenfunctions and eigenvalues.   As discussed in Section~\ref{subsec:CD-FPCA}, under IC conditions the CD-FPCA scores $\xi_{lr}$ obtained through~\eqref{eq:generate_scores} satisfy $\mathbb{E}(\xi_{lr}\mid z_l)=0$ and $\mathrm{Var}(\xi_{lr}\mid z_l)=\nu_r(z_l)$ with approximately uncorrelated components. Consequently, raw scores are not directly comparable across covariate levels, because their dispersion changes systematically with the environmental state. In our application, however, the model-based eigenvalue curves $\hat{\nu}_r(z)$ were not sufficiently stable across the covariate range and did not consistently match the empirical dispersion of the estimated scores in \eqref{eq:generate_scores}. We therefore introduced a post-FPCA calibration step based on the Phase~I scores. For each retained component $r$, we estimated a smooth mean curve $\hat m_r(z)$ using a generalized additive model (GAM; \citealp{Wood_2017}) and subtracted it from the scores as a covariate-dependent mean correction. The squared centered scores were then fitted by a GAM with log link, yielding a positive variance estimate $\hat v_r(z)$. The standardized monitoring inputs were thus defined as
\begin{equation}\label{eq:score_standardization}
    X_{lr}=\frac{\hat{\xi}_{lr}-\hat m_r(z_l)}{\sqrt{\hat v_r(z_l)}},\qquad r=1,\ldots,m,
\end{equation}
and collected in $\boldsymbol{X}_l=(X_{l1},\ldots,X_{lm})^\top$ for the subsequent application in a multivariate control chart.

\subsection{MEWMA control chart on standardized scores}
To detect small-to-moderate persistent changes, the Multivariate Exponentially Weighted Moving Average (MEWMA) chart of \citet{Lowry.etal_1992} to the standardized score vectors $\{\boldsymbol{X}_l\}$ is applied. The MEWMA state is defined recursively as
\begin{equation}\label{eq:mewma_state_rewrite}
\boldsymbol{Z}_l = \lambda \boldsymbol{X}_l + (1-\lambda)\boldsymbol{Z}_{l-1},
\qquad \boldsymbol{Z}_0=\boldsymbol{0},\qquad 0<\lambda\le 1,
\end{equation}
where smaller values of $\lambda$ yield stronger smoothing and hence greater sensitivity to sustained deviations.
Under ideal in-control conditions with $\mathrm{Cov}(\boldsymbol{X}_l)=\boldsymbol{I}_m$, the stationary covariance of $\boldsymbol{Z}_l$ is
$\boldsymbol{\Lambda}_{\boldsymbol{Z}}=\frac{\lambda}{2-\lambda}\,\boldsymbol{I}_m$,    
which leads to the monitoring statistic
\begin{equation}\label{eq:mewma_T2_rewrite}
T_l^2 = \boldsymbol{Z}_l^\top \boldsymbol{\Lambda}_{\boldsymbol{Z}}^{-1}\boldsymbol{Z}_l = \frac{2-\lambda}{\lambda}\,\boldsymbol{Z}_l^\top \boldsymbol{Z}_l.
\end{equation}
An alarm is raised whenever $T_l^2 > h_4$, where $h_4$ is selected to attain a prescribed in-control ``average run length'' ARL$_0$.
Because the standardized Phase~I scores could still exhibit mild residual serial dependence, we calibrated $h_4$ using a moving-block bootstrap, as in \cite{Neumann.etal_2025c}, rather than relying exclusively on the i.i.d. Gaussian approximation. Bootstrap sequences are generated by repeatedly sampling contiguous blocks of consecutive daily score vectors $\boldsymbol{X}_i$ of length $b$ from Phase I data. For the control limit $h_4$, the run length was defined as $\mathrm{RL}(h_4)=\inf\{i\geq 1:T_i^2>h_4\},$ and the in-control ARL was estimated by averaging this run length over 100,000 bootstrap replicates. The final control limit was then obtained numerically by utilizing a grid search to match the desired target ARL.
\section{Modeling and Monitoring Results on the KW51 Extended Data}\label{sec:results}
The complete post-retrofitting period comprised 1101 days. The observed Mode 13 frequency profiles were strongly incomplete, only 17 days contained complete 24-point trajectories, while 1084 days exhibited at least one missing profile observation. On average, 16.5 frequency observations per day were available, ranging from 0 to 24, which corresponds to an overall missingness rate of 31.2\%. Temperature measurements were also incomplete, but less sparse overall, with an average of 19.0 observations per day. Between 18th May 2021 and 27th May 2021, maintenance works were carried out on the KW51 railway bridge. During this period, the northern rails of tracks A and B were replaced. The minimum daily temperature spanned a fairly narrow mild-temperature band between 8.7°C and 11.2°C. The new rails were delivered on 18th May 2021 and temporarily stored on the sleepers next to the existing rails to be renewed. This temporary storage corresponded to an additional distributed mass of approximately $2 \times 60\,\text{kg/m}$ along the bridge and may therefore have affected its dynamic behavior. In the night from 23rd May 2021 to 24th May 2021, the outer rail of track A was renewed. In the night from 26th May 2021 to 27th May 2021, the outer rail of track B was renewed, and the old rails for both tracks were removed. 
In the present study, Phase I extended from 27th May 2019, immediately after completion of the retrofitting, to 31st March 2021, whereas Phase II covered the period from 1st April 2021 to 30th September 2022. The maintenance period lies within Phase II and serves as the event window for evaluating the monitoring results obtained with both the ‘old’ CAFDA-SHM framework and the new CD-FPCA-based approach.

\subsection{Temperature effects on the mean and (co-)variance}\label{subsec:estimated_basic_model}
\begin{figure}[h]
    \centering
    \includegraphics[width=1\linewidth]{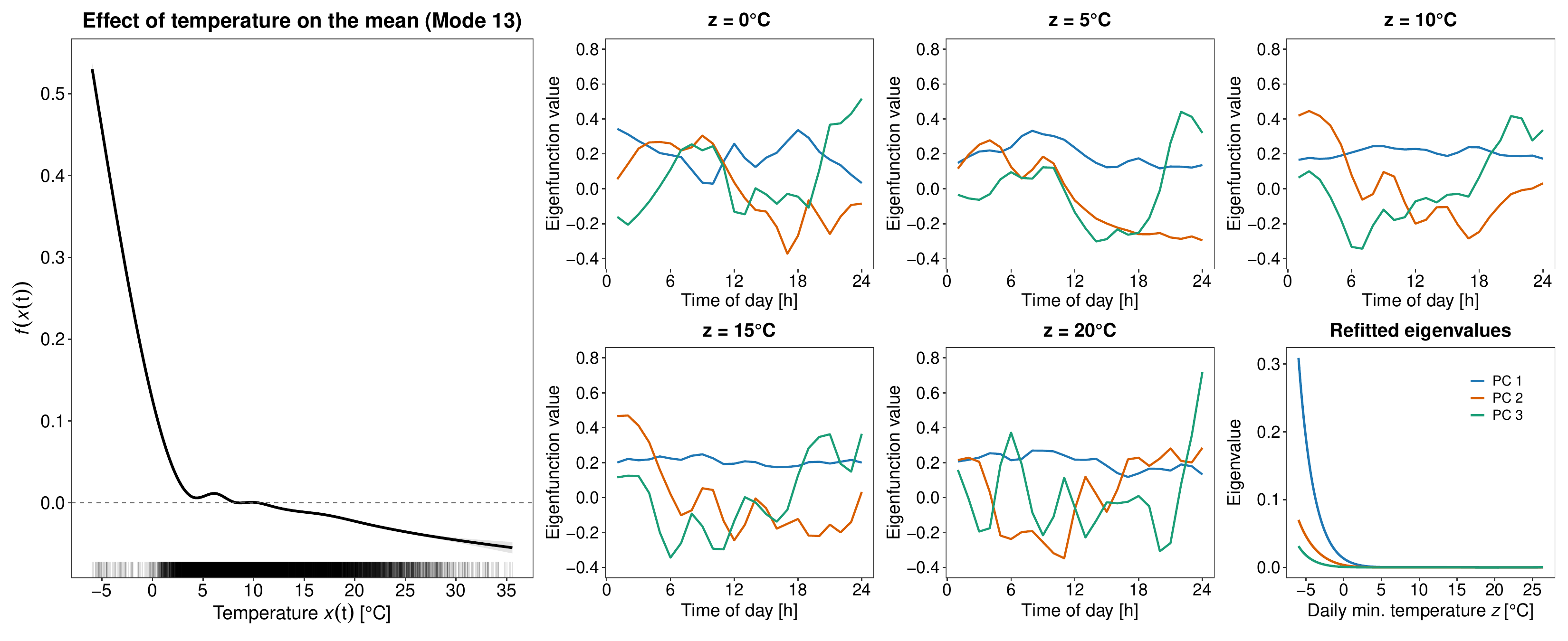}
    \caption{Temperature effect on the mean $\mu$ (left), covariate-dependent eigenfunctions (middle) and eigenvalues (right) for various temperatures.}   \label{fig:estimated_eigenfunctions_eigenvalues}
\end{figure}

Figure~\ref{fig:estimated_eigenfunctions_eigenvalues}~(left) shows the estimated (fixed) covariate effect $f\!\left(x(t)\right)$ on the mean $\mu$. The functional intercept (not shown) is nearly flat, indicating that after accounting for temperature, (almost) no additional systematic time-of-day structure remains in the mean response. However, the estimated temperature effect is very strong and clearly nonlinear, particularly in the lower temperature range. This suggests that temperature is the dominant driver of the expected frequency profile, with more severe changes occurring under cold conditions. These lower-temperature regimes are of particular importance for the subsequent monitoring analysis, since they are sparsely represented in the data and are associated with increased instability and environmentally induced alarms. Overall, the fixed-effects model explains approximately $R^2 \approx 0.80$ of the observed variation.

The remaining plots in Figure~\ref{fig:estimated_eigenfunctions_eigenvalues} show, as an example, the first three eigenfunctions for five distinct temperatures between 0°C and 20°C, along with the corresponding eigenvalues as smooth functions of the daily minimum temperature. The eigenvalues, which correspond to the variances of the respective (functional) principal component scores, decrease monotonically with increasing temperature. The eigenfunctions also vary with temperature but appear rather wiggly, making them more difficult to interpret. At least, the first component (blue) consistently captures the overall level of the residual process because it is basically constant across both temperature and time.   

\subsection{Monitoring results}
For both control charts (using either CAFDA or CD-FPCA), we set $\lambda=0.4$ and targeted an in-control average run length of ARL$_0=500$, which is well aligned with the scale of the available reference period and implies approximately one expected false alarm over an in-control run of similar length. In addition, both charts used the same fixed-effects model as described in Subsection~\ref{subsec:estimated_basic_model}.

\begin{figure}[h]
    \centering
    \includegraphics[width=1.0\linewidth]{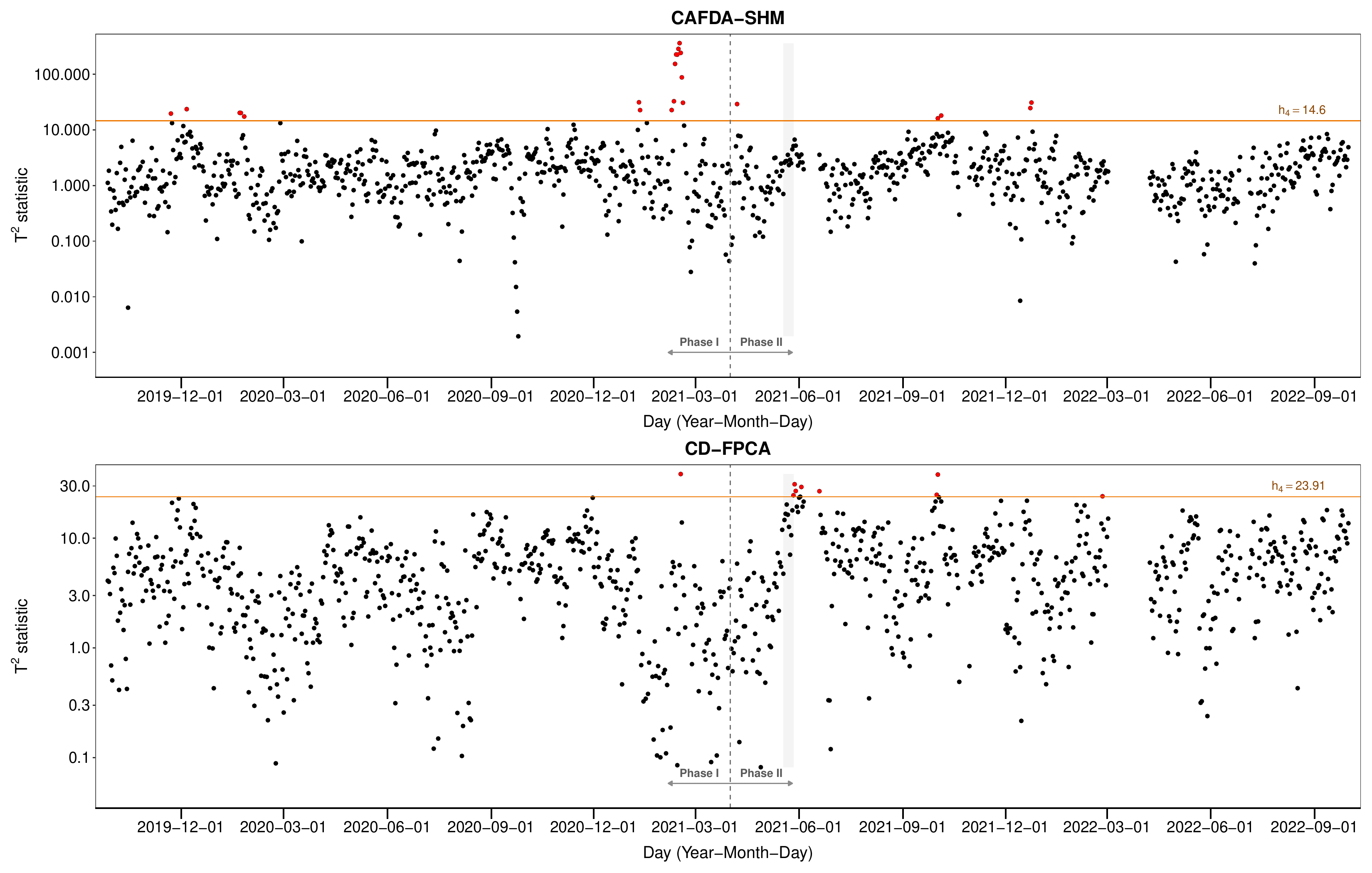}
    \caption{MEWMA control charts for the CAFDA-SHM (top) and the proposed CD-FPCA framework (bottom).}
    \label{fig:method_comparison}
\end{figure}

For the CAFDA-SHM, the control limit was determined under the i.i.d.\ assumption as in \citet{Wittenberg.etal_2025a}. The estimated number of functional principal components was $m=3$, corresponding to a fraction of explained variance of 95\%, and the resulting alarm threshold was $h_4=14.6$. Using the full temperature spectrum without additional robustification in Phase~I, the CAFDA-SHM chart exhibited a pronounced cold-temperature issue in February 2021 as displayed in Figure \ref{fig:method_comparison}~(top). In total, 17 alarms were observed in Phase~I, of which 14 occurred on days with temperatures at or below 2°C. 
In Phase~II, the pattern was more mixed but still showed a substantial temperature-related component. A total of 5 alarms were recorded, with 3 occurring near freezing temperatures. The remaining 2 alarms occurred in October under comparatively mild conditions of around 12°C. At the same time, the chart did not detect the bridge intervention event highlighted in grey in Figure~\ref{fig:method_comparison}.
For the CD-FPCA approach, the control limit was set to $h_4=23.91$. Its calibration required a moving-block bootstrap with block length $b=3$, due to the weak but non-negligible serial dependence observed in the empirical autocorrelation function of the standardized scores in \eqref{eq:score_standardization}. The resulting MEWMA chart is shown in Figure~\ref{fig:method_comparison}~(bottom). In Phase~I, the chart signaled only a single false alarm on 16th February 2021 (at around 2.3°C). In Phase~II, eight signals were detected. The first alarm occurred on 27th May 2021, i.e., on the last day of the bridge intervention event, and was followed by two further signals on the next two consecutive days. The remaining signals, on 3rd June, 19th June, 1st and 2nd October in 2021, and on 25th February 2022 under comparatively milder temperatures, are likely false alarms. Compared with CAFDA-SHM, this indicates a substantial reduction of environmentally induced (false) alarms in Phase~I while still yielding a detection during the event window.
\section{Discussion and Outlook}\label{sec:discussion_conclusion}
The case study from Section~\ref{sec:results} represented a challenging real-world setting. The tracked natural-frequency profiles are sparsely and irregularly observed, exhibit substantial missingness, and the anomaly associated with the bridge intervention is comparatively small. In addition, low-temperature regimes, which are particularly relevant for the covariance structure, are less densely represented. These features make the dataset informative for assessing the robustness of the proposed approach, but they also limit the extent to which a sharp separation between in-control and out-of-control behavior can be expected. Analyses in the pre-retrofitting setting, not included in this paper, indicate a stronger performance in the presence of a substantially larger shift during the retrofitting period. Thus, the monitoring performance observed here should be interpreted in light of the particular difficulties of the (new) post-retrofitting case study.

Further, the proposed CD-FPCA framework has certain limitations and suggests avenues for future research. First, the new method was implemented only for a scalar covariate, the minimum daily temperature. An important extension would therefore be a fully functional covariate-dependent FPCA that accounts for the complete temperature profile rather than a scalar summary. Second, the present analysis employs a comparatively simple functional mean model and does not incorporate further environmental or operational variables. The proposed approach was also applied only to a single system output (Mode 13), although \citet{Wittenberg.etal_2025b} showed that related monitoring concepts can be extended to multivariate functional data streams. Finally, instead of the local-linear three-dimensional smoother used here, alternative estimators such as the spline-based covariate-dependent FPCA approach of \citet{Ding.etal_2022} could be explored in future work.
\section*{Software}
The statistical analyses were conducted in the open-source software \texttt{R} \citep{R_2025}. Functional additive (mixed) models were fitted with the \texttt{mgcv} package \citep{Wood_2017}. In addition, the CD-FPCA algorithm, originally implemented in MATLAB \citep{MATLAB_2022b}, was ported to \texttt{Julia} \citep{Bezanson.etal_2017} and used for the corresponding analyses.
\section*{Acknowledgements}
This research paper out of the project `SHM -- Digitalisierung und Überwachung von Infrastrukturbauwerken' is funded by dtec.bw -- Digitalization and Technology Research Center of the Bundeswehr, which we gratefully acknowledge. dtec.bw is funded by the European Union -- NextGenerationEU.

\bibliographystyle{abbrvnat}
\bibliography{literature}
\end{document}